\documentclass[twocolumn,aps,prl,reprint]{revtex4-1}
\usepackage{blindtext}
\usepackage{graphicx}
\usepackage{amsmath}
\usepackage[margin = 0.7in]{geometry}
\usepackage{dcolumn}
\usepackage{tabularx}
\usepackage{gensymb}
\usepackage{xcolor}
\colorlet{ins}{blue} 
\colorlet{del}{red}
\usepackage{ulem}
\usepackage{nameref}

\begin{document}
\title{Anomalous frequency and temperature dependent scattering in the dilute metallic phase in lightly doped-SrTiO$_3$}
\author{K. Santhosh Kumar$^1$}
\author{Dooyong Lee$^2$,$^3$}
\author{Shivasheesh Varshney$^2$}
\author{Bharat Jalan$^2$}
\author{N. P. Armitage$^1$}
\email{npa@jhu.edu}
\affiliation{$^1$Institute for Quantum Matter and Department of Physics and Astronomy, John Hopkins University, Baltimore, Maryland-21218, USA.}
\affiliation{$^2$Department of Chemical Engineering and Materials Science, University of Minnesota, Minneapolis, Minnesota 55455, USA.}
\affiliation{$^3$Department of Physics Education, Kyungpook National University, 80, Daehak-ro, Buk-gu, Daegu 41566, South Korea.}

\begin{abstract}
The mechanism of superconductivity in materials with aborted ferroelectricity and the emergence of a dilute metallic phase in systems like doped-SrTiO$_3$ are outstanding issues in condensed matter physics. This dilute metal has features both similar and different to those found in the normal state of other unconventional superconductors.  We have investigated the optical properties of the dilute metallic phase in doped-SrTiO$_3$ using THz time-domain spectroscopy. At low frequencies the THz response exhibits a Drude-like form as expected for typical metal-like conductivity.  We observed the frequency and temperature dependencies to the low energy scattering rate $\Gamma(\omega, T) \propto (\hbar\omega)^2 + (p \pi k_BT)^2 $ expected in a conventional Fermi liquid, despite the fact that densities are too small to allow current decay through electron-electron scattering.   However we find the lowest known $p$ values of 0.39-0.72. As $p$ is 2 in a canonical Fermi liquid and existing models based on energy dependent elastic scattering bound $p$ from below to 1, our observation lies outside current explanation.  Our data also gives insight into the high temperature regime and shows that the temperature dependence of the resistivity derives in part from strong T dependent mass renormalizations.
\end{abstract}

\maketitle

Superconductivity in doped-SrTiO$_3$ is intriguing for a number of reasons.  It onsets at carrier concentrations as low as $\sim$ 5.5 $\times$ 10$^{17}$/cm$^3$ \cite{gastiasoro2020superconductivity, lin2013fermi} , shows a coexistence of superconductivity and ferroelectric fluctuations \cite{scheerer2020ferroelectricity}, and exhibits a dome-like shape of T$_c$ as a function of carrier concentration reminiscent of the high-T$_c$ superconducting cuprates (albeit one with a much lower maximum T$_c \sim$ 0.5 K \cite{scheerer2020ferroelectricity, rischau2022isotope}).  Quantum oscillation measurements and tunneling experiments suggest an unusual two-band superconductivity that is consistent with the expectation for the electronic structure at this doping \cite{binnig1980two, lin2014critical}.  The very small Fermi energy does not allow the usual condition ($E_F \ge \omega_{D}$) for the retarded interaction of the BCS and Migdal-Eliashberg theories and cannot explain the pairing mechanism in SrTiO$_3$ \cite{marsiglio2020eliashberg}. Several physical mechanisms have been proposed in the literature to explain the superconductivity in SrTiO$_3$ including the exchange of two optical phonons \cite{appel1969soft, yu2022theory, van2019possible, wolfle2018superconductivity, ngai1974two}, plasmons \cite{ruhman2016superconductivity}, dynamically screened electron-longitudinal optical (LO) phonon interactions \cite{klimin2012microscopic}, ferroelectric quantum fluctuations \cite{edge2015quantum}, and Cohen's theory of many-valley semiconductors \cite{cohen1964superconductivity}. In this respect, the pairing mechanism in the superconducting state of SrTiO$_3$ remains unresolved.

Moreover, the unusual transport properties of the normal state of doped-SrTiO$_3$ distinguishes it from  several other doped semiconductors. For example, it has a robust metallic phase that persists into the extremely dilute limit (likely owing to its large Bohr radius \cite{collignon2019metallicity}) and a quadratic temperature dependence of electrical resistivity $\rho (T) = \rho_0 +AT^2$ (where $\rho_0$ is the residual resistivity) despite its small Fermi surface \cite{behnia2022origin}.  Such Fermi-liquid like $T^2$-scaling of resistivity is widely observed in many strongly correlated electron systems, but aspects of this scaling here are anomalous.  The ``$A$" coefficient exhibits a modified Kadowaki-Woods scaling (observed in many correlated metals) \cite{behnia2022origin, jacko2009unified} in which $A \propto E_F^{-2}$.  The doping dependence of the pre-factor and its magnitude is large \cite{lin2015scalable} at low dopings and approaches that of heavy-fermion compounds.  However, the fact that Fermi surface is in general too small to engage in umklapp scattering and only a single band is occupied at the lowest dopings where T$^2$ is also observed (which makes the interband Baber scattering inapplicable) makes it an open question how $T^2$ resistivity can arise from electron-electron scattering in this material.
There are a few proposals that make connection to the low temperature superconductivity such as an electron-soft phonon interaction and phonon or disorder mediated electron-electron interactions \cite{appel1969soft, yu2022theory, van2019possible, wolfle2018superconductivity,klimin2012microscopic}.  However, similar $T^2$-dependence has been observed in another single-band dilute superconductor's normal state Bi$_2$O$_2$Se, which has no soft phonon \cite{wang2020t}.  The $T^2$ resistivity dependence switches to other T dependencies at higher temperature and does not show saturation even up to $\sim$ 1000 K \cite{collignon2019metallicity,collignon2020heavy}. Therefore, it is a ``bad metal" in that the metallic resistivity exceeds the Mott-Ioffe-Regel limit. Despite the extensive studies, the microscopic mechanism for current decay and the critical role of the soft phonon mode in $T^2$ resistivity in the metallic state \cite{zhou2018electron,zhou2019predicting}, and its relation to superconductivity remain open questions.

\begin{figure*}
    \centering
    \includegraphics[width = 18cm]{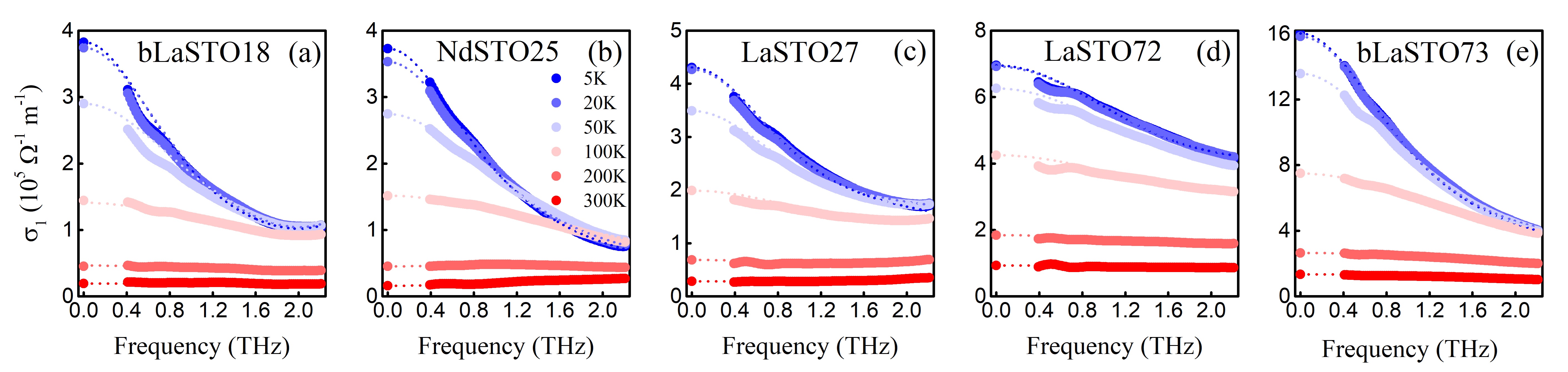}
    \caption{The temperature dependent real part of the THz complex conductivity of doped-SrTiO$_3$ thin films as a function of frequency at various carrier concentration (a) bLaSTO18 (1.8 $\times$ 10$^{20}$/cm$^3$, 100 nm thickness), (b) NdSTO25 (2.5 $\times$ 10$^{20}$/cm$^3$, 194 nm thickness), (c) LaSTO27 (2.7 $\times$ 10$^{20}$/cm$^3$, 100 nm thickness), (d) LaSTO72 (7.2 $\times$ 10$^{20}$/cm$^3$, 66 nm thickness), and (e) bLASTO73 (7.3 $\times$ 10$^{20}$/cm$^3$, 100 nm thickness). DC conductivity obtained using four probe measurement shown at zero frequency. Dotted lines indicate the model fittings as discussed in the text.}
    \label{Fig.1}
\end{figure*}

To understand the unusual temperature dependent dc electrical resistivity in the dilute metallic phase of doped-SrTiO$_3$, we performed time-domain THz (TDTS) spectroscopy measurements on doped-SrTiO$_3$ thin films. The THz complex conductivity approximately follows the functional form for the Drude model as expected for simple metallic conduction. The power of these experiments is they allow one to seperate out the effects of spectral weight changes (proportional to $n/m$) and scattering.  The effective optical mass ($m^*$) from Drude model fitting is nearly constant at low temperature, but shows an unusual enhancement at higher temperature. The scattering rate scales as T$^2$ below 120 K and increases slower at temperatures above this range. To investigate the possibilities for Fermi liquid-like scattering, we employed an extended Drude model analysis that reveals an $\omega^2$ dependent scattering rate. We observed the standard quadratic scaling of temperature and frequency of the optical scattering rate, albeit one with a unconventionally small size of the temperature dependent contribution as compared to the energy dependent contribution. We find relative values below that of any known model.  Our data also shows that the temperature dependence of the resistivity at high temperature derives in part from strong T dependent mass.

Epitaxial La-and Nd-doped SrTiO$_3$ thin films were grown on (001) oriented (LaAlO$_3$)$_{0.3}$(Sr$_2$AlTaO$_6$)$_{0.7}$ (LSAT) single-crystal substrates using hybrid molecular beam epitaxy (EVO-50 Scienta-Omicron). Films were grown under stoichiometric conditions with different doping densities which resulted in different carrier concentrations and disorder levels (as characterized by the residual resistivity ratios (RRR)) \cite{yue2022anomalous}.  bLaSTO73 (7.3 $\times$ 10$^{20}$/cm$^3$; RRR = 20) and bLaSTO18 (1.8 $\times$ 10$^{20}$/cm$^3$; RRR = 20) were grown with a SrTiO$_3$ buffer layer whereas the LaSTO72 (7.2 $\times$ 10$^{20}$/cm$^3$; RRR = 6), LaSTO27 (2.7 $\times$ 10$^{20}$/cm$^3$; RRR = 14), and NdSTO25 (2.5 $\times$ 10$^{20}$/cm$^3$; RRR = 12) were grown directly on the LSAT(001) substrate. This gave a range of carrier concentrations that span 1.79-7.29 $\times$ 10$^{20}$/cm$^3$.  Further growth details of the films are provided in the Supplementary Material (SM).  Hall resistivity shows a weak dependence on temperature indicating that it is likely is a measure of the carrier concentration (See SM). Now standard TDTS experiments were performed as described in the SM \cite{krewer2018accurate}.  

Fig.~1 shows the real part of the THz optical conductivity of the doped-SrTiO$_3$ thin films in increasing order of carrier concentrations. Corresponding imaginary ($\sigma_2(\omega)$) parts at various temperatures are shown in Fig.~S2(a). At low temperature, $\sigma_1(\omega)$ falls off with increasing frequency, behavior which is expected for typical metal-like conduction.  We also observe a high frequency tail in $\sigma_1(\omega)$ that is more prominent at low dopings. We believe this is a remnant of the transverse optical phonon soft mode (TO1) that is relevant for the paraelectric state in undoped SrTiO$_3$.  As the temperature increases, carrier scattering increases and $\sigma_1(\omega)$ decreases with a broadened Lorentzian that is consistent with an increase in dc resistivity. Notably also the area of the conductivity peaks appear to diminish as T increases.   The same basic behavior is observed in the series NdSTO25, LaSTO27, LaSTO72, and bLaSTO73 with increasing carrier concentration. The soft phonon mode appears to shift to high energies with increasing carrier concentration and its contribution to the high concentration films becomes negligible as observed in LaSTO72 and bLaSTO73 \cite{crandles1999optical,van2008electron}. Importantly as temperature increases the integrated spectral weight of all curves decreases (Figs. S3 and S4).  We will relate this below to an increasing mass at higher temperature.

\begin{figure}{b}
    \centering
    \includegraphics[width = 8.6cm]{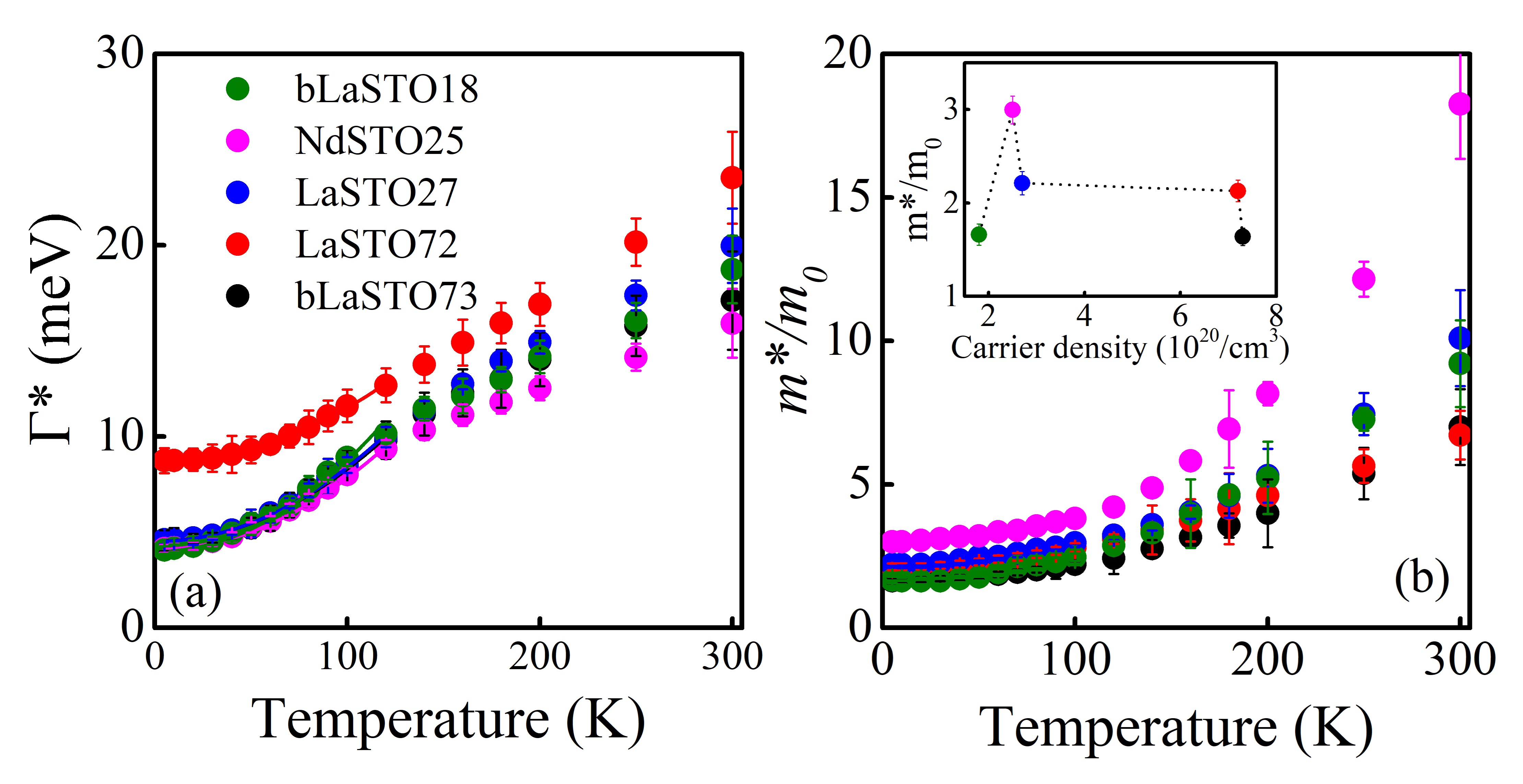}
    \caption{The temperature dependent Drude model fitting parameters in doped-SrTiO$_3$ thin films at various carrier densities (a) scattering rate ($\Gamma^*$) and (b) effective mass ($m^*/m_0$). Solid lines in (a) indicate $T^2$ fitting to the scattering rate.  Mass is given in units of the free electron mass ($m_0$). Carrier density dependent mass at 5 K is provided in the inset of (b).}
    \label{Fig.2}
\end{figure}

To further evaluate the conductivity quantitatively, we fit it to a model with a zero frequency Drude term plus a higher energy Lorenz oscillator to account for the residual soft mode and a background dielectric constant \cite{kumar2020terahertz,kumar2022unraveling}.

\begin{equation}
    \sigma(\omega) =  \frac{\omega_{p}^{*2} \varepsilon_{0}}{\Gamma^*-i\omega } +\frac{S^{2} \omega \varepsilon_{0} \gamma}{\omega+i (\omega_0^2-\omega^2) \gamma}-i\omega \varepsilon_0 (\varepsilon_{\infty}-1)
\end{equation}
where $\omega_{p}^{*2}$ is the spectral weight of the Drude peak (equivalent to the plasma frequency squared renormalized by effective mass effects e.g. $\omega^{*2}_p = \omega^{2}_p \frac{m_b}{m^*} =  \frac{ne^2}{m_b \varepsilon_0} \frac{m_b}{m^*} )$, $\Gamma^*$ is the renormalized Drude scattering rate (also renormalized by the effective mass factor), $S^2$ is the spectral weight of the soft phonon, $\gamma$ is the width of the soft phonon, $\omega_0$ is the frequency of the soft phonon, and $i\omega \varepsilon_0 (\varepsilon_{\infty}-1)$ is the background lattice contribution. Here, the soft phonon mode frequencies are taken from Ref. \cite{marsik2016terahertz}.

\begin{figure*}{t}
    \centering
    \includegraphics[width = 18cm]{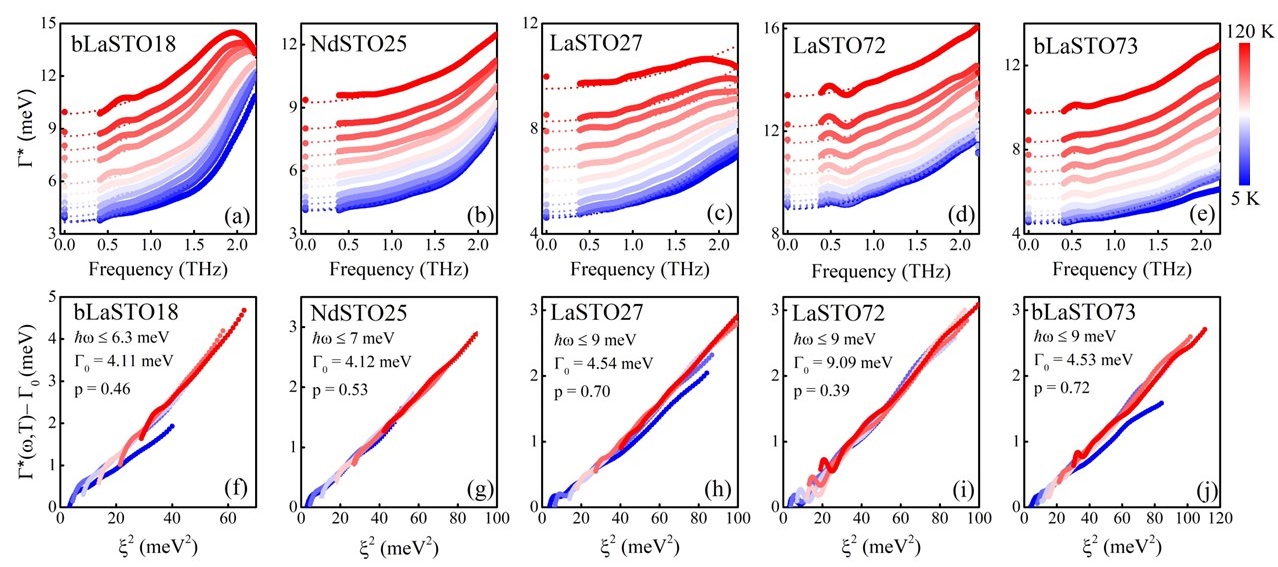}
    \caption{Frequency dependent scattering rate of the doped-SrTiO$_3$ thin films at various carrier concentration shown in (a) bLaSTO18, (b) NdSTO25, (c) LaSTO27, (d) LaSTO72, and (e) bLASTO73 in the temperature range of 5(blue)-120 K(red). Dotted lines indicate a $\omega^2$ fitting to the experimental data. The data points at zero frequency are the Lorentzian Drude scattering widths shown in Fig.~2(a).  The Gurzhi scaling for the collapse of the relaxation rate $\Gamma^*(\omega, T)$ in doped-SrTiO$_3$ thin films (f) bLaSTO18, (g) NdSTO25, (h) LaSTO27, (i) LaSTO72, and (j) bLASTO73. The corresponding $\hbar \omega$, $\Gamma_0$, and $p$ value are given in each plot. A T$_{max}$ of 50 K was used for all samples in the analysis.}
    \label{Fig.3}
\end{figure*}

The Drude scattering rate ($\Gamma^*$) obtained from the fittings is plotted in Fig.~2(a).  Larger error bars near room temperature are due to the flat optical conductivity.
It is found that the optical scattering rate closely follows a quadratic temperature dependence, $\Gamma^*  = \Gamma_0 +A'T^2$ ($\Gamma_0$ is the impurity contribution) below 120 K where the dc electrical resistivity does the same. We plot the extracted $A'$ from fitting as a function of density in Fig.~4(a).  $A'$ decreases with increasing carrier density with a similar dependence as observed for the $A$ coefficient in dc transport measurements \cite{lin2015scalable}.

The effective masses ($m^*/m_0$) were extracted using the relation $\omega^{*2}_p$ = $ne^2/m^* \varepsilon_0$, where the carrier density ($n$) is obtained from Hall transport measurements, $e$ is the electron charge, and $\omega^{*2}_p$ is the spectral weight (proportional to the area) of the Drude peak. As shown in Fig.~2(b), $m^*/m_0$ is 1.6 - 3 times the free electron mass for all samples at 5 K and only weakly temperature dependent up to 100 K. Above 100 K, it increases faster with increasing temperature. 
The temperature dependence of the $m^*/m_0$ enhancement is generally stronger for low carrier densities. Such behavior was previously observed and interpreted as due to strong electron-LO phonon coupling where the depleted Drude spectral weight [Fig.~S3] appeared at the LO phonon mode around ~250 meV \cite{van2008electron}.  Such free electron-like conduction with an enhanced effective mass at high temperature indicates Fröhlich type large polaron conduction \cite{zhou2018electron,zhou2019predicting,chen2015observation,verdi2017origin}.  Our data shows that the temperature dependence of the dc electrical resistivity at higher temperatures ($\rho = m^*/ne^2\tau^*$; $\Gamma^* = 1/\tau^*$)  originates via a combination of the temperature dependence of $\Gamma^*$ and $m^*/m_0$.  In contrast, at the lowest temperatures the effective mass remains nearly constant and the dc electrical resistivity goes as $T^2$ following the T$^2$-dependence of $\Gamma^*$. 

Fermi liquid-like behaviour is inferred from the T$^2$ dependence of the dc resistivity below 120 K.  In order to measure the frequency dependent scattering rate in doped-SrTiO$_3$, we analyzed the THz conductivity spectrum using the extended Drude model \cite{kumar2020terahertz}.  One can analyze the frequency dependent scattering rate from the expression $  \Gamma(\omega) = \omega_{p}^{2} \varepsilon_{0} \mathrm{Re} \frac{1}{\sigma(\omega) } $.  Here $\omega_p$ is the plasma frequency from all intraband spectral weight (in principle determined from integrating the conductivity to high energies after subtracting the phonon contribution) and $\varepsilon_{0}$ is the free space dielectric constant.  However, to make the connection to the fitted Drude widths above, we plot instead in Fig. 3(a-e) the renormalized scattering rate, which is $\Gamma^*(\omega) =  \frac{ \omega^{2}_p  }{ \omega^{*2}_p  } \Gamma (\omega)$.   Note that via the definition of $\Gamma$ above the undetermined factor of $\omega^{2}_p $ drops out in $\Gamma^*$.  $\Gamma^*(\omega)$ is found to increase with frequency over the entire doping range. The strong frequency dependence in $\Gamma^*(\omega)$ above 1.5 THz at low carrier concentration is likely evidence for a soft phonon contribution to the conductivity. Further, the frequency dependence of $\Gamma^*(\omega)$ stays the same up to room temperature for bLaSTO73 and LaSTO72.  The optical scattering rate follows a quadratic frequency dependence $\Gamma^*(\omega) =  \Gamma_0 + b \omega^2$, where $\Gamma_0$ is due to impurities and defects, and $b$ is a prefactor. The obtained $b$ from the fitting is plotted as a function of carrier density in Fig.~4(a) (using same fitting range as for the scaling plots in Fig. 3).

A standard check for Fermi liquid behaviour is the collapse of inelastic optical relaxation rate according to the Gurzhi scaling law \cite{stricker2014optical,pustogow2021rise}
\begin{equation}
\Gamma(\omega, T) - \Gamma_0 =  b\omega^2 +  A' T^2  \propto \xi^2 \equiv (\hbar\omega)^2 + (p \pi k_BT)^2,
\end{equation}
where $p$ is the Gurzhi scaling factor that is expected to be 2 for a canonical Fermi liquid [29]. The Gurzhi scaling plots for $\Gamma^*(\omega, T)$ as function of $\xi^2$ shown in Fig.~3(f-j) for doped-SrTiO$_3$ thin films at various carrier concentrations.
Although $T^2$ dependence of the quasi-particle scattering rate is observed up to 120 K, in this analysis we consider only the temperature range below 50 K (which is the range where the Drude spectral weight and effective mass remains nearly constant).  Although the Gurzhi scaling is expected to apply to $\Gamma$ not $\Gamma^*$ we can apply it to $\Gamma^*$ because of the weak T dependence of the mass below 50 K. We also considered energies only below some cutoff (indicated in the plots), above which the frequency dependence changed (presumably due to the soft phonon). The calculated $p$ as function of temperature for doped-SrTiO$_3$ thin films at various carrier concentrations is plotted in Fig.~4(a). Another way to calculate the scaling parameter is $A'/b$ = $(p\pi)^2$ where $A'$ and $b$ are the pre-factors of the temperature and frequency dependent scattering rates, respectively. The calculated $p$ values approximately matches those obtained from the scaling plots. We find the $p$ values are in the range from 0.39-0.72 as shown in Fig.~4(b). One can note from Fig.~4(a) that $A$, $A'$, and $b$ of LaSTO72 deviates from carrier density trend of other samples. Similarly, $p$ is also reduced for LaSTO72 which has $\Gamma_0$ larger than other samples. It likely indicates that $p$ has a dependence on $\Gamma_0$ (impurity scattering contribution). Generally $p$ increases with increasing carrier density as shown in  Fig.~4(b).

The quadratic temperature and frequency dependent scattering rate, and dc electrical resistivity suggest a Fermi liquid state in the dilute metallic doped-SrTiO$_3$, however, with a small $p$ of 0.39-0.72. The deviation of $p$ from 2 is common in many strongly correlated materials including transition metal oxides, organic metals, and heavy Fermion materials [Please see Table S1 in SM] \cite{stricker2014optical,pustogow2021rise,yang2006temperature,katsufuji1999frequency,nagel2012optical,sulewski1988far,mirzaei2013spectroscopic}, where energy dependent elastic scattering contributes to the energy dependent but not to the temperature dependent quasi-particle self-energy \cite{maslov2012first} [36]. However, such models give a lower bound on $p$ of 1, and so our observation is anomalous and needs explanation.

It is important to reiterate that these observations are being made in low density system where $\omega^2$ scattering is already anomalous as the Fermi wave vector is too small at concentrations below 2 $\times$ 10$^{21}$/cm$^3$ to allow umklapp scattering.  The three bands occupation in our doping range means that in principle interband Baber scattering is a possible source of dissipation. However the fact that $T^2$-resistivity has been observed down to ranges where only a single band is occupied (carrier concentration below 1.2$\times$ 10$^{18}/cm^3$) does not -- in our estimattion -- make it a likely source of the effect. Although sources of scattering other than electronic are possible, we will  reiterate that the $T^2$ resistivity coefficient appeared to show a modified Kadowaki-Woods scaling \cite{behnia2022origin,jacko2009unified} in which $A \propto E_F^{-2}$ that would seem to confirm the dominance of essentially electronic mechanisms.  However, it is possible that electron-phonon coupling plays a role. In general, the negligible net polarization by the soft TO phonon mode does not scatter electrons. However, this is only true for a spherical Fermi surface. Electron scattering from a resonant soft mode could be possible in a non-spherical Fermi surface of doped-SrTiO$_3$ \cite{ehrenreich1956scattering,wemple1966evidence}. Polaron physics may also be relevant as dopings are close to the regime where the Fermi energy exceeds the LO4 phonon energy \cite{chen2015observation,verdi2017origin}.  Since the electron-phonon coupling is very strong in doped-SrTiO$_3$, theoretical studies likely need to further consider phonon mediated electron-electron interactions \cite{macdonald1980electron}.

In this work, we have investigated the THz complex conductivity response of the doped-SrTiO$_3$ thin films using THz time-domain spectroscopy. The THz complex conductivity in doped-SrTiO$_3$ thin films is similar to the Drude model conductivity which is expected for metal-like conduction. However, although the carrier effective mass obtained from the spectral weight is nearly constant at low temperature it shows unusual enhancement with increasing temperature.  The scattering rate scales as T$^2$ below 120 K.  Our results further indicate that the temperature dependence in the dc resistivity at high temperature is due to the combined effect of both effective mass and scattering rate changes. At low temperatures an extended Drude model analysis of the THz conductivity reveals an $\omega^2$ dependence of scattering rate. This quadratic dependence of $\Gamma^*(\omega)$ follows the Fermi liquid like Gurzhi scaling for $\Gamma(T, \omega)$ in doped-SrTiO$_3$, however, with a relative scale of the T$^2$ dependence below that of any known model. Its likely that an electronic scattering mechanism dominates the low temperature resistance, albeit one that is aided by electron-phonon interactions.  It is likely that understanding these aspects of the normal state interactions are important for understanding the appearance of the anomalous superconductivity.

\begin{figure}
    \centering
    \includegraphics[width = 8.9cm]{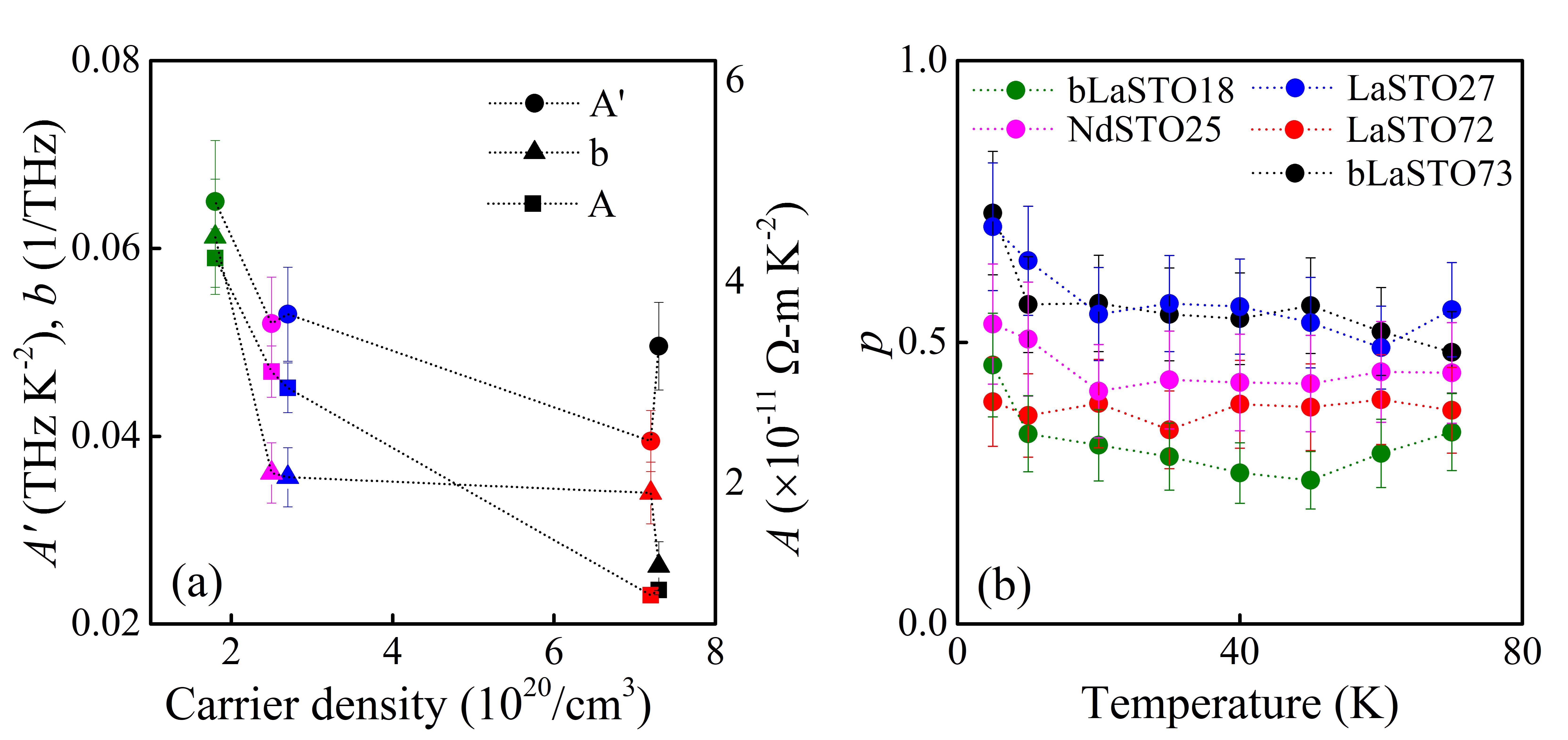}
    \caption{(a) The carrier density dependent pre-factors, $A'$ is the T$^2$ scattering rate pre-factor, $b$ is the $\omega^2$ scattering rate pre-factor, and $A$ is the T$^2$ dc electrical resistivity pre-factor. (b) Shows the temperature dependent Gurzhi scaling parameter ($p$) for doped-SrTiO$_3$ thin films.}
    \label{Fig.4}
\end{figure}

The project at JHU was supported by the NSF-DMR2226666 and the Gordon and Betty Moore Foundation’s EPiQS Initiative through Grant No. GBMF9454. NPA had additional support by the Quantum Materials program at the Canadian Institute for Advanced Research. Synthesis and electrical transport (S.V. and B.J.) were supported by the U.S. Department of Energy through DE-SC0020211, and in part by the Center for Programmable Energy Catalysis, an Energy Frontier Research Center funded by the U.S. Department of Energy, Office of Science, Basic Energy Sciences at the University of Minnesota, under Award No. DE- SC0023464.  D.L. acknowledge support from the Air Force Office of Scientific Research (AFOSR) through Grant Nos. FA9550-21-1-0025. Parts of this work were carried out at the Characterization Facility, University of Minnesota, which receives partial support from the NSF through the MRSEC program under award DMR- 2011401. Film growth was performed using instrumentation funded by AFOSR DURIP award FA9550-18-1-0294.

\bibliography{main}


\end{document}